\documentclass[12pt, preprint]{aastex}
\begin{document}

\title{Rapid Transition of Uncombed Penumbrae to Faculae during Large Flares}
\author{Wang, Haimin\altaffilmark{1}, Deng, Na\altaffilmark{2,1} and Liu, Chang\altaffilmark{1}}
\affil{1. Space Weather Research Laboratory, New Jersey
Institute of Technology, Newark, NJ 07102, haimin.wang@njit.edu}
\affil{2. Department of Physics and Astronomy, California State University, Northridge, CA, 91330}

\begin{abstract}

In the past two decades, the complex nature of sunspots has been disclosed with high-resolution observations. One of the most important findings is the ``uncombed'' penumbral structure, where a more horizontal magnetic component carrying most of Evershed Flows is embedded in a more vertical magnetic background \citep{Solanki+Montavon1993A&A...275..283S}. The penumbral bright grains are locations of hot upflows and dark fibrils are locations of horizontal flows that are guided by nearly horizontal magnetic field. On the other hand, it was found that flares may change the topology of sunspots in $\delta$ configuration: the structure at the flaring polarity inversion line becomes darkened while sections of peripheral penumbrae may disappear quickly and permanently associated with flares \citep{LiuC+etal2005ApJ...622..722L}. The high spatial and temporal resolution observations obtained with Hinode/SOT provide an excellent opportunity to study the evolution of penumbral fine structure associated with major flares. Taking advantage of two near-limb events, we found that in sections of peripheral penumbrae swept by flare ribbons, the dark fibrils completely disappear, while the bright grains evolve into faculae that are signatures of vertical magnetic flux tubes. The corresponding magnetic fluxes measured in the decaying penumbrae show stepwise changes temporally correlated with the flares. These observations suggest that the horizontal magnetic field component of the penumbra could be straightened upward (i.e., turning from horizontal to vertical) due to magnetic field restructuring associated with flares, which results in the transition of penumbrae to faculae.

\end{abstract}


\section{INTRODUCTION}

Solar flares have been understood as the result of magnetic reconnection in the solar corona \citep[see recent review by][]{Hudson2011SSRv..158....5H}.
Although magnetic field evolution in solar photosphere plays important roles in building energy and triggering eruption, most models of flares imply that photospheric magnetic fields do not have rapid, irreversible changes associated with the eruptions. The traditional picture is that the solar surface, where the coronal magnetic fields are anchored, has much higher density and gas pressure than the corona. In recent years, however, rapid and irreversible changes of photospheric magnetic field have been found to be closely associated with flares \citep[e.g.,][]{WangH1992SoPh..140...85W, Wang+etal1994ApJ...424..436W, Kosovichev+Zharkova2001ApJ...550L.105K, Spirock+etal2002ApJ...572.1072S, Wang+etal2002ApJ...576..497W, Yurchyshyn+etal2004ApJ...605..546Y, Wang+etal2004ApJ...605..931W, Sudol+Harvey2005ApJ...635..647S, WangH2006ApJ...649..490W, Wang+Liu2010ApJ...716L.195W, Petrie+Sudol2010ApJ...724.1218P, WangS+etal2011arXiv1103.0027W}. A flare-associated change of sunspot white-light structure was also discovered
\citep{WangH+etal2004ApJ...601L.195W, Deng+etal2005ApJ...623.1195D, LiuC+etal2005ApJ...622..722L, ChenW+etal2007ChJAA...7..733C}. I.e., Pieces of peripheral sunspot penumbra can suddenly disappear right after a flare. In particular, \citet{LiuC+etal2005ApJ...622..722L} discussed the peripheral penumbral decay and central penumbral darkening of $\delta$ spots seen in a number of X-class flares, and proposed a reconnection picture where the central (i.e., near flaring polarity inversion line (PIL)) magnetic fields collapse inward resulting in a more horizontal configuration and the peripheral fields turn toward the flaring PIL resulting in a more vertical configuration after flares. These observational results have been summarized recently by \citet{Wang+Liu2010ApJ...716L.195W} in the context of the conjecture of \citet{Hudson+Fisher+Welsch2008ASPC..383..221H} and \citet{Fisher+etal2010arXiv1006.5247F}. These authors quantitatively assessed the back reaction on the solar surface and interior resulting from drastic coronal field evolution required to release energy, and made the prediction that after flares, the coronal magnetic fields should collapse toward the flaring PIL. In addition, flare induced sudden lateral motions of penumbrae were also found \citep{Gosain09,Mattews11}.

On a different front, sunspot penumbrae have been suggested to exhibit an ``uncombed'' structure \citep{Solanki+Montavon1993A&A...275..283S}. The basic idea is that the penumbral fields have two intermingled magnetic components: a more vertical magnetic background called ``spines'' and a more horizontal magnetic component called ``intraspines'' \citep{lites+etal1993}. There is a broad consensus that the intraspines carry most of Evershed flows throughout the entire penumbra \citep[e.g.,][]{BellotRubio+etal2003A&A...403L..47B, BellotRubio+etal2004A&A...427..319B, Borrero+etal2005A&A...436..333B, Ichimoto+etal2007PASJ...59S.593I, Borrero+Solanki2008ApJ...687..668B, Deng+etal2010ApJ...719..385D}. Reflecting in photospheric intensity images, the intraspines are manifested as (1) bright grains or filaments (a.k.a. bright heads) in the inner and middle penumbra representing locations of hot Evershed upflows \citep[e.g.,][]{rimmele+marino2006, Ichimoto2010mcia.conf..186I}, and (2) dark fibrils (a.k.a. dark tails) following those bright grains/filaments and extending to the middle and outer penumbra corresponding to cooled horizontal Evershed flows that are guided by nearly horizontal magnetic field \citep[e.g.,][]{stanchfield+thomas+lites1997, schlichenmaier+jahn+schmidt1998b, Rempel2011ApJ...729....5R}. These are evidenced by recent observations and simulations that the Evershed flows measured by Dopplergrams correlate with brighter features in the inner penumbra and with darker features in the outer penumbra \citep{Schlichenmaier+etal2005AN....326..301S, BellotRubio+etal2006A&A...453.1117B, Ichimoto+etal2007PASJ...59S.593I, Rempel2011ApJ...729....5R}. It should be noted that in the outer penumbra, the spines could appear relatively brighter than the horizontal-flow-carrying intraspines seen as dark fibrils. Thus, in short, the bright grains or filaments in the inner and middle penumbra are most likely associated with the inner portion of intraspines, while bright filaments in the outer penumbra may also be associated with spines.  It is also worth mentioning that the outer ends of intraspines especially in the inner penumbra do not necessarily dip into the photosphere but may tilt upward \citep[see Figure 19 of][]{Rempel2011ApJ...729....5R}, and that intraspines in the outer penumbra are more likely to turn back into the photosphere. The uncombed penumbral structure has been corroborated by recent advanced space- and ground-based spectro-polarimetric observations as well as radiative magnetohydrodynamic (MHD) modeling \citep[e.g.,][]{Ichimoto+etal2007PASJ...59S.593I,Rempel+etal2009Sci...325..171R, Jurcak+BellotRubio2008A&A...481L..17J,BeckC2011A&A...525A.133B,Rempel2011ApJ...729....5R,Borrero+Ichimoto2011LRSP....8....4B}.

We note that besides the uncombed penumbra model, ``gappy penumbra model'' \citep{Spruit+Scharmer2006A&A...447..343S, Scharmer+Spruit2006A&A...460..605S} explains the penumbra in a different scenario for high-resolution observations in the order of 0.1$^{\prime\prime}$, in which the dark cores in penumbral grains can be resolved. In this model, the penumbral filamentary structure is interpreted as due to convection in radially aligned field-free gaps beneath a nearly potential magnetic field. The gappy model can also explain many aspects of penumbral structure however, at present the angular resolution of the spectropolarimetric observations is insufficient to verify it.

To facilitate our analysis and discussion, we first present an investigation on the uncombed penumbral structure (Section~\ref{structure}), and then base the interpretation of our observations (Section~\ref{flare}) on the framework of the uncombed penumbra model depicted using schematic cartoons (Section~\ref{discussion}). We mainly concern ourselves with a natural question that when sunspot penumbra suddenly disappears, what would happen with respect to the uncombed magnetic structure? The high-resolution observation from Hinode covering major flares provides an excellent opportunity to answer this question. We demonstrate evidences that at the site of the penumbral decay, dark fibrils completely disappear while bright grains/filaments evolve into faculae that are signatures of vertical magnetic flux tubes. These observations suggest that the intraspines are straightened upward to become more vertical. Therefore, the original uncombed magnetic structure seems to be re-organized toward the vertical direction as a consequence of the field restructuring due to flares, which results in the disappearance of penumbrae.

\section{OBSERVATIONS AND RESULTS}

\subsection{The ``Umcombed'' Penumbral Structure} \label{structure}
We study the relationship between intensity and magnetic inclination of the detailed penumbral structure using Hinode Spectro-Polarimeter (SP) vector magnetograms\footnote{http://sot.lmsal.com/data/sot/level2d/} obtained for a $\beta$ sunspot near disk center in the NOAA Active Region (AR) 10923 on 2006 November 14 07:15:04~UT, the results of which are presented in Figure~1. The vector data were retrieved using High Altitude Observatory Milne-Eddington inversion \citep[e.g.,][]{Lites+etal2008ApJ...672.1237L}, and we further transformed the observed fields to the local Cartesian coordinates so that the inclination angle could be measured with respect to the local surface normal (i.e., 0$^{\circ}$ is vertical and 90$^{\circ}$ is horizontal). We applied the threshold method around the median value to individually identify the bright and dark penumbral fibrils (enclosed by the yellow and purple contours, respectively, in Figure~1, left column). Besides analyzing the overall intensity and magnetic inclination distribution for the entire penumbra (Figure~1, middle column), we also trace the shape of the spot (red contours in Figure~1, left column) to examine the azimuthal averages of intensity and inclination angle at different normalized radius (Figure~1, right column). The results of histograms and radial profiles unambiguously indicate that the dark penumbral fibrils generally exhibit a more horizontal orientation than the bright fibrils, with larger difference nearer the outer penumbra. This is consistent with the results derived based on the ground-based high-spatial resolution observations of line-of-sight (LOS) magnetograms \citep{Langhans+etal2005A&A...436.1087L}, and is thus in line with the uncombed penumbral model with two distinct but related magnetic components. We note that in the uncombed model, the dark fibrils that we identified could be the outer portion of intraspines, while the bright fibrils could be either the inner portion of intraspines or spines.

\subsection{Rapid Restructuring of Penumbrae during Flares} \label{flare}
The high spatial and temporal resolution observations obtained with Hinode/Solar Optical Telescope \citep[SOT;][]{Tsuneta+etal2008SoPh..249..167T} on 2006 December~6 and 2007 June~4 provide useful data sets to study the evolution of penumbral fine structure associated with major flares.
An X6.5 flare occurred at 18:29~UT on 2006 December 6 in AR 10930 (S06E60) and an M8.9 flare occurred at 05:06~UT on 2007 June 4 in AR 10960 (S06E51). The main data source is the continuous G-band images that have a nominal cadence of 1 or 2 minutes and pixel scale of 0.11$^{\prime\prime}$ covering the flares.

Figure 2 shows selected images exhibiting the rapid disappearance of a penumbral section (marked by the red box) associated with the X6.5 flare on 2006 December 6. The lower row images clearly show the transitional decaying phases of the penumbra within an hour after the onset of the flare. As reference, a stable penumbral area and a facular area are selected and marked by green and yellow boxes, respectively. Figure 3 shows images before and after the M8.9 flare on 2007 June 4. Two mpeg movies for the two events are provided as online material to depict more dynamical detail. Here are the few points that we noted by studying the time sequence of images.

(1) Morphologically, before the flares, the rapidly changing areas appear as regular penumbral structure, consisting of bright grains/filaments and dark fibrils lying at the periphery of sunspots. After the flares, the morphology is similar to the area consisting of faculae.

(2) As these two regions are far away from disk center, the faculae
(G-band bright points) appear much more obviously comparing to the regions close to disk center, such as on 2006 December 13 when an X3.4 flare occurred \citep{Tan+etal2009ApJ...690.1820T}. This is likely due to the ``hot wall''
effect --- the faculae contrast enhances when ${\rm cos}\theta$ is between 0.5 and 0.1,
where $\theta$ is the heliocentric angle \citep{spruit1976}. This gives us the opportunity to observe the structure and evolution of bright points that correspond to magnetic elements. Unfortunately, high-resolution magnetograms from SOT/SP were not available to cover these two flares. Therefore, we use the faculae as a proxy for vertical flux tubes.

(3) Based on the analysis of movies, it is clear that the bright points appear in-situ immediately after the penumbral decay, i.e., they were not transported from other areas.

(4) We can not trace all the bright points continuously due to data gaps, but it is obvious that some penumbral bright grains/filaments evolve into faculae. Statistically, the number of post-flare faculae is similar to that of pre-flare penumbral bright grains/filaments in the same area. For example, we identified 10 local maxima in the form of penumbral bright grains/filaments inside the red box before the flare on 2006 December 6. 10 local maxima are also identified in the form of faculae in the same box after the flare. For the 2007 June 4 flare, 8 local maxima are identified in the form of penumbral bright grains/filaments before the flare, while 9 local maxima are identified in the form of faculae after the flare.

(5) The penumbrae started to decay immediately after the flare brightening swept across them. Within an hour after the onset of the flare, the penumbral dark fibrils gradually lose their filamentary structure and progressively disappear from the outer part toward the inner part, which leads to a reduction of penumbral area and an increase of the corresponding brightness. Finally, the dark fibrils completely disappear, leaving bright facular points at the locations where the original penumbral bright grains/filaments are located.

(6) After the penumbrae decayed, the area of the parent umbrae apparently increased.

To quantitatively demonstrate the evolution of the decaying penumbrae, in Figures 4 and 5 we construct the time sequence of intensity histogram for the outlined areas shown in Figures 2 and 3. Each figure has three panels. The center panel is for the rapidly decaying penumbral area. The left and the right panels are for the referential stable penumbral and facular regions, respectively. In each panel, the X-axis marks the time and the Y-axis marks the normalized intensity. The colors along the Y-direction represent the intensity distribution (histogram) of all pixels inside each outlined area, i.e., the brightness represents the fraction of those pixels falling in the individual intensity range. Soft X-ray and hard X-ray flare light curves are overplotted. It is obvious that only the decaying penumbral areas have a sudden change of the intensity distribution closely associated with the flares. Specifically, their intensity distribution in the pre-flare state well resembles that of the regular penumbra, while that in the post-flare state becomes like the facular region. The change of the penumbral structure associated with the X6.5 flare is more abrupt (noticeable changes took place within a few minutes and reached the stable level in about a half hour), while the intensity distribution of the 2007 June 4 region associated with the M8.9 flare evolves gradually for about 40 minutes before the data gap.

To further explore the change of the magnetic structure, in Figures 6 and 7 we use LOS magnetograms from the Michelson Doppler Imager (MDI) and Global Oscillation Network Group (GONG) to study the magnetic flux change in the decaying penumbral areas. It can be clearly seen that for both events, the flux has a rapid stepwise change temporally correlated with the flares, with a significant decreases for AR 10930 while a pronounced increases for AR 10960. We draw schematic pictures to explain the different behavior of the flux change in these two active regions close to the limb. For the X6.5 flare occurred in AR 10930, the decaying penumbra is located at the disk-ward side of the flaring PIL, in which case the flux would decrease as the field lines turn from horizontal to vertical (i.e., away from the LOS direction). On the contrary, for the M8.9 flare occurred in AR 10960, the decaying penumbra is located at the limb-ward side of the flaring PIL, in which case the flux would increase when the field lines turn from horizontal to vertical (i.e., toward the LOS direction). Therefore, the rapid flux variation in the two events supports a common change of the field lines turning from horizontal to vertical direction in the peripheral, decaying penumbral region.

\section{SUMMARY AND DISCUSSION} \label{discussion}

Based on the G-band observations presented above, we conclude that the rapid decay of penumbra has two distinct aspects: the dark penumbral fibrils disappear completely, while bright penumbral grains/filaments evolve to faculae. The magnetic fluxes measured in the decaying penumbrae show stepwise changes temporally correlated with the flares, suggesting that those magnetic fields turn from horizontal to vertical.

To accommodate the observation of penumbral decay in the uncombed penumbral model, we propose a scenario as illustrated in Figure 8 highlighting the changes of field structure in the penumbra region associated with the flare occurrence. The upper panel depicts a $\delta$ spot in the preflare state with uncombed penumbral structure as described in Section~1, which consists of spines (blue lines) and intraspines (red lines). A majority of Evershed flows (red arrows) are carried by the intraspines. The bright grains/filaments (yellow spots) indicate the inner portion of intraspines in the inner and middle penumbra or spines in the outer penumbra. In the postflare state as shown in the lower panel of Figure 8, the horizontal segment of intraspines (the black shaded portion) could turn vertical hence leading to the disappearance of dark fibrils. At the locations of the original bright penumbral grains/filaments manifesting the inner portion (footpoints) of the intraspines, the straightened flux tubes naturally appear as faculae subsequently. Meanwhile, the spines may also become more vertical following the overall trend, so that the corresponding bright filaments evolve to faculae as well. As a result, the original uncombed magnetic structure of penumbra seems to be combed toward the vertical direction as the consequence of flares, which results in the transition of penumbrae to faculae. For the field lines near umbra, they may merge into the umbral fields as they become more vertical, which can then explain the observed increase of the umbral area.

We further discuss the observed rapid changes of penumbral structure and the possible interpretation and implication as follows.

(1) In a $\delta$ configuration (see Figure 8), part of the penumbrae is located at the central region in between the two umbrae with opposite polarities and usually exhibits highly sheared configuration along the PIL, while other portions of penumbrae lie at the periphery of the $\delta$ spot and is separated from the PIL by its parent umbra. We must clarify that all the rapid penumbral decays associated with flares that we have found so far \citep[][and in this paper]{WangH+etal2004ApJ...601L.195W, Deng+etal2005ApJ...623.1195D, LiuC+etal2005ApJ...622..722L, ChenW+etal2007ChJAA...7..733C} occur at the peripheral penumbra region. As a comparison, we found that the penumbrae at the central region are usually enhanced after flares (i.e., the intensity becomes lower and the horizontal Evershed flow increases) \citep[see e.g.,][]{LiuC+etal2005ApJ...622..722L, DengN+etal2011ApJ...733L..14D}, even though they may also be swept by flare ribbons. This also explains that in the cases studied by Li \& Zhang (2009), the sudden penmubral decay was not observed while the flare ribbons swept sunspots.

(2) In order that the horizontal segment of the intraspine flux tubes could easily turn vertical, the dark fibrils carrying horizontal Evershed flows may be relatively shallow while bright grains may be deeply anchored. This is supported by observations of helioseismic and spectropolarimetric inversions \citep{gizon+duvall+larsen2000, Borrero+etal2006A&A...450..383B, BeckC2011A&A...525A.133B}. The magneto-convection with similar properties exist in both quiet region and penumbra in the forms of granulation and anisotropic granulation, respectively \citep{Rempel2011ApJ...729....5R}. This might also be supported by the finding of \citet{KuboM+etal2008ApJ...681.1677K} that when the penumbral boundary moves inward, the granules appear in the outer boundary. The observation of sudden disappearance of penumbra in the 2006 December 13 X3.4 event close to disk center demonstrates the similar pattern, in which the regular granules appear immediately after the penumbra disappear \citep{Tan+etal2009ApJ...690.1820T}.

(3) The rapid penumbral decay is temporally and spatially correlated with flares, which indicates the crucial role of the field restructuring due to flares in such a rapid decay process different from the general sunspot evolution. Although the decayed penumbrae are swept across by flare ribbons, the penumbral decay is not due to transient heating or brightening induced by flares, as the decay is permanent and non-reversible indicating that the original uncombed magnetic fields must undergo some fundamental changes most probably toward the vertical direction. Although we believe that it is the flare-induced field change that leads to the permanent penumbral decay, it is not clear whether the flare heating plays a role in the mass flow within penumbral flux tubes to help the horizontal magnetic fields to turn vertical more easily.


(4) As mentioned previously, the peripheral penumbral decay could be accompanied by the enhancement (darkening) of the central penumbrae \citep[e.g.,][]{LiuC+etal2005ApJ...622..722L, DengN+etal2011ApJ...733L..14D} as the field lines at the flaring PIL become more horizontal after flares \citep{Wang+Liu2010ApJ...716L.195W}, which is consistent with the ``implosion'' picture of \citet{Hudson2000ApJ...531L..75H} \citep[also see][]{LiuR+WangH2009ApJ...703L..23L}. Recently, this was further discussed in terms of back reaction of coronal field restructuring on the solar surface, as the implosion would result in a downward Lorentz force \citep{Hudson+Fisher+Welsch2008ASPC..383..221H, Fisher+etal2010arXiv1006.5247F}. We speculate that when the magnetic energy in the flaring core region is decreased, a negative pressure core would form above the flaring PIL attracting ambient fields to fill the void. Subsequently, the central magnetic fields collapse downward to become more horizontal near the PIL, while the fields in the peripheral region turn more vertical toward the central region (see Figure 8). As reflected in white-light observations, the central penumbrae could then be enhanced (darkened), and portions of the peripheral penumbrae may decay depending on the flare magnitude, the distance of the peripheral penumbrae to the PIL, and whether being swept by flare ribbons (i.e., in direct association with the flaring process). The analysis of the three-dimensional MHD simulations of an emerging twisted flux tube \citep{FanY2010ApJ...719..728F} also evidences the enhanced and more inclined field at the central flaring PIL together with the decreased and more vertical peripheral fields after the eruption and the associated flares \citep{LiY+etal2011ApJ...727L..19L}. As related, \citet{KuboM+etal2011ApJ...731...84K} elaborated that the decay of penumbra must be closely associated with the increase of field inclination.

(5) As \citet{BellotRubio+etal2008ApJ...676..698B} observed, sunspot penumbra may decay slowly in the course of a few days leaving the structure of ``naked'' sunspots (i.e., spots without penumbrae; \citealt{Liggett+Zirin1983SoPh...84....3L}). When dark fibrils disappear, some bright structures (they name as ``fingers'') appear and carry upward flows in the order of 1--2~km~s$^{-1}$. The ``fingers'' may be similar flux tubes as the faculae discussed in the current paper. In this sense, flares may accelerate the decay process of sunspot penumbra in the outskirt to within an hour instead of days. Interestingly, \citet{RempelM2011SPD....42.1001R} found that the top boundary conditions determine the radial extent of penumbra in the simulated sunspots, and that the penumbra forms or destructs within 0.5 hours after the top boundary changes. This is very similar to our observations where the penumbra disappears rapidly within an hour after the onset of the flares, which suggest a reconfiguration of coronal fields accompanied with violent changes in the upper atmosphere that resembles the change of the top boundary in the simulation. Therefore, our observations strongly support the possible impact by the coronal transients on the photospheric magnetic structure of sunspots.

\acknowledgments

The authors thank Drs. K. Ichimoto, M. Rempel, and R. Liu for reading the manuscript and very helpful comments to improve the paper. We also thank the referee for valuable comments that help us to improve the paper, and Yuan Yuan for help on image processing. Hinode is a Japanese mission developed and launched by ISAS/JAXA, with NAOJ as domestic partner and NASA and STFC (UK) as international partners. It is operated by these agencies in co-operation with ESA and NSC (Norway).
This work is supported by NSF grants AGS-0839216, AGS-0819662, and AGS-0849453, NASA grants NNX08AQ90G, NNX11AQ55G and NNX11AO70G. ND is supported by NASA grant NNX08AQ32G.

\begin{figure}
\epsscale{1.0}
\plotone{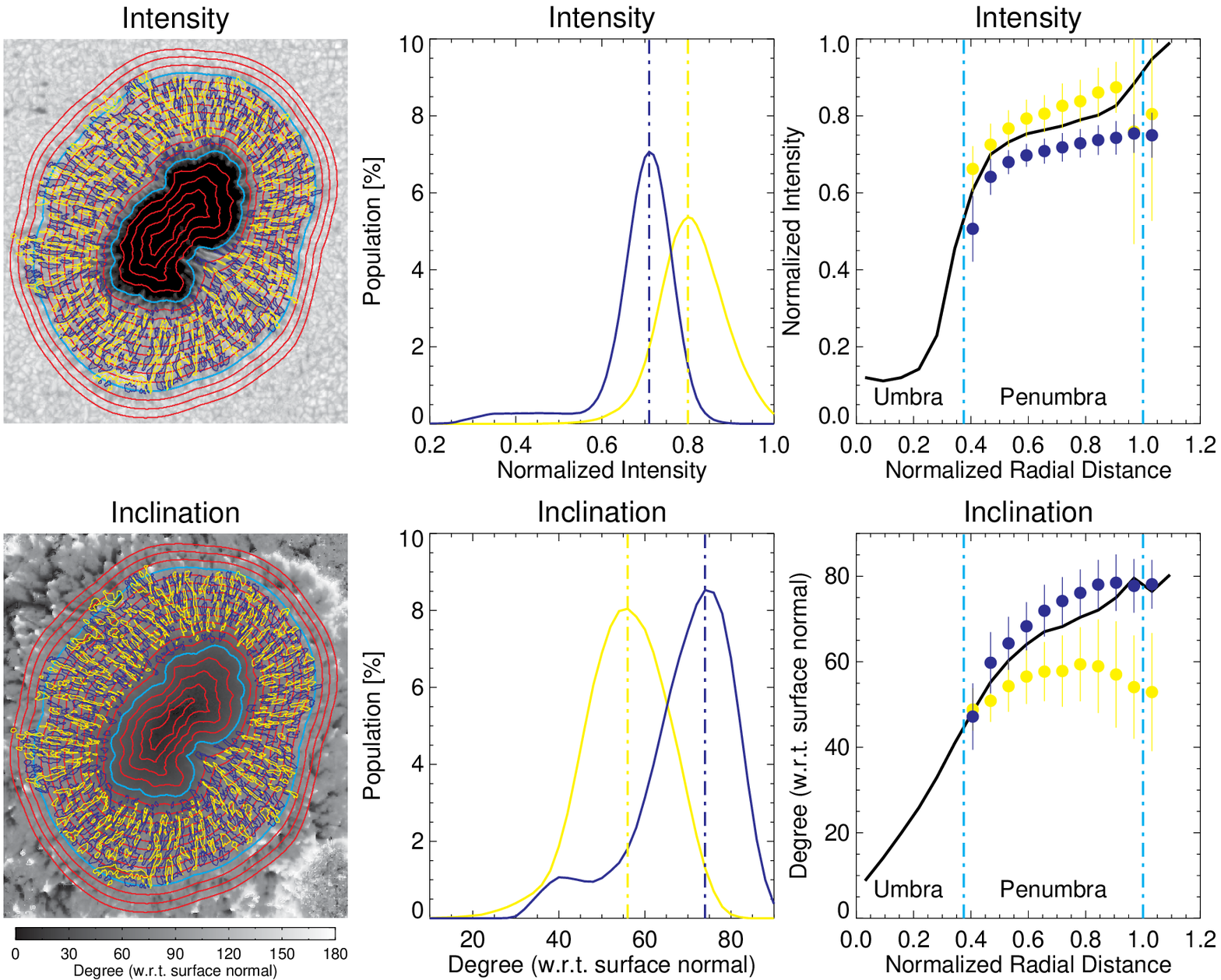}
\caption{Left column: Hinode observation of the continuum intensity and magnetic inclination of a $\beta$ spot in the NOAA AR 10923 near disk center. The inclination angle is measured with respect to the local surface normal (i.e., 0$^{\circ}$ is vertical and 90$^{\circ}$ is horizontal). The yellow and purple contours mask the penumbral bright and dark fibrils, respectively. The blue contours outline the inner and outer boundaries of the penumbra. The red contours following the shape of the spot are used to calculate the azimuthal averages of intensities and inclination angles at different radius. Middle column: Intensity and inclination histograms of the bright (yellow) and dark (purple) fibrils in the entire penumbra. Right column: The intensity and inclination angle of the bright (yellow) and dark (purple) fibrils averaged azimuthally at each normalized radial distance (0 is the sunspot center, 1.0 indicates the outer boundary of penumbra). The standard deviations of each azimuthal average are plotted as error bars. The black lines are the azimuthal averages over all pixels at each normalized distance.}
\end{figure}

\begin{figure}
\epsscale{1.0}
\plotone{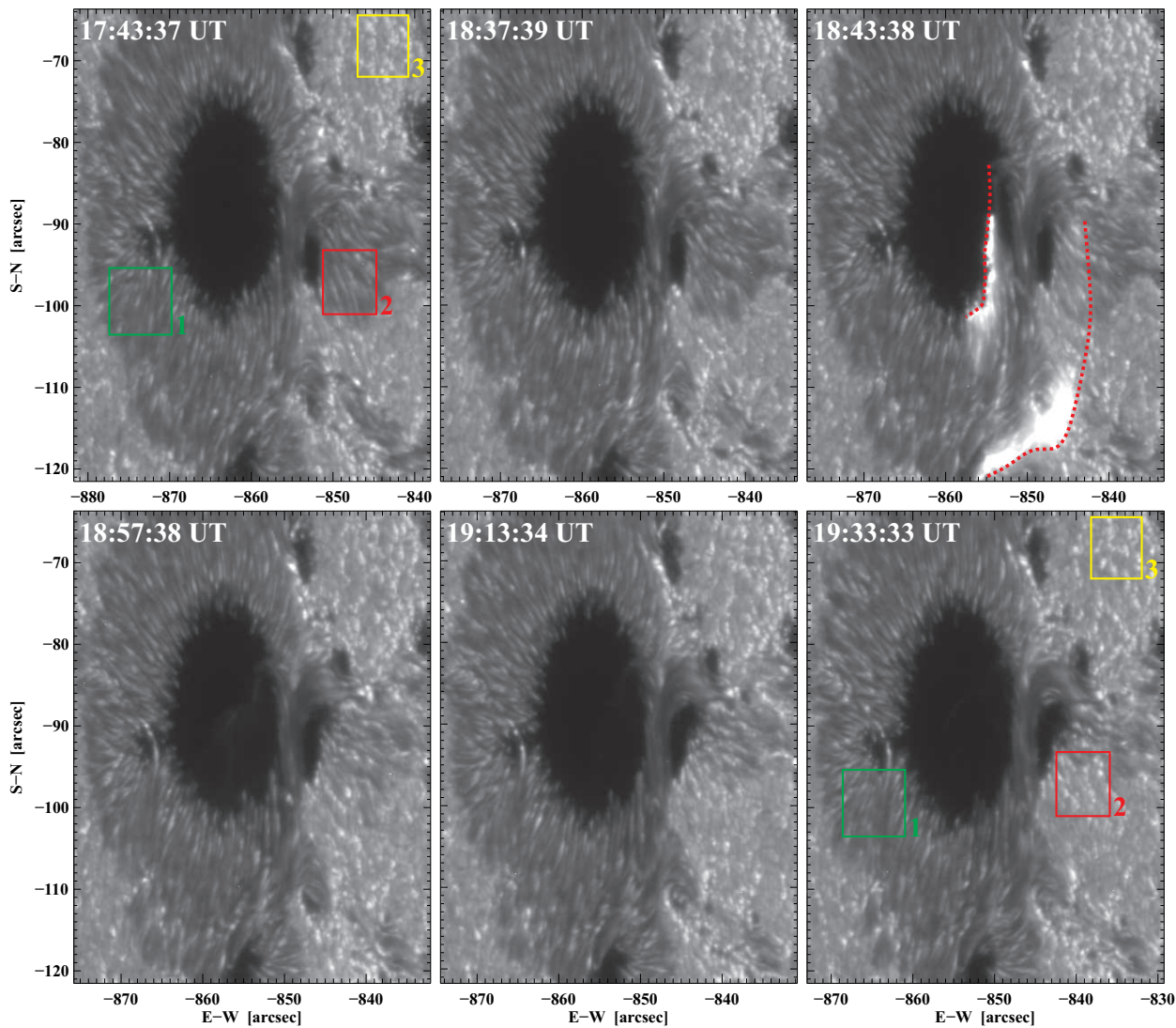}
\caption{Time sequence of G-band images observed by Hinode/SOT on 2006 December 6 covering the X6.5 flare. The red box marks the region of a decaying penumbra (in the entire active region 10930, there are several more penumbral decay regions; see the accompanied online movie and Figure 2 of \citealt{DengN+etal2011ApJ...733L..14D}). The green and yellow boxes mark the referential stable penumbral and facular region, respectively. The red dotted lines in the frame of 18:43:38 UT delineate the separating flare ribbons, with the western one sweeping across the penumbral decay region.}
\end{figure}

\begin{figure}
\epsscale{1.0}
\plotone{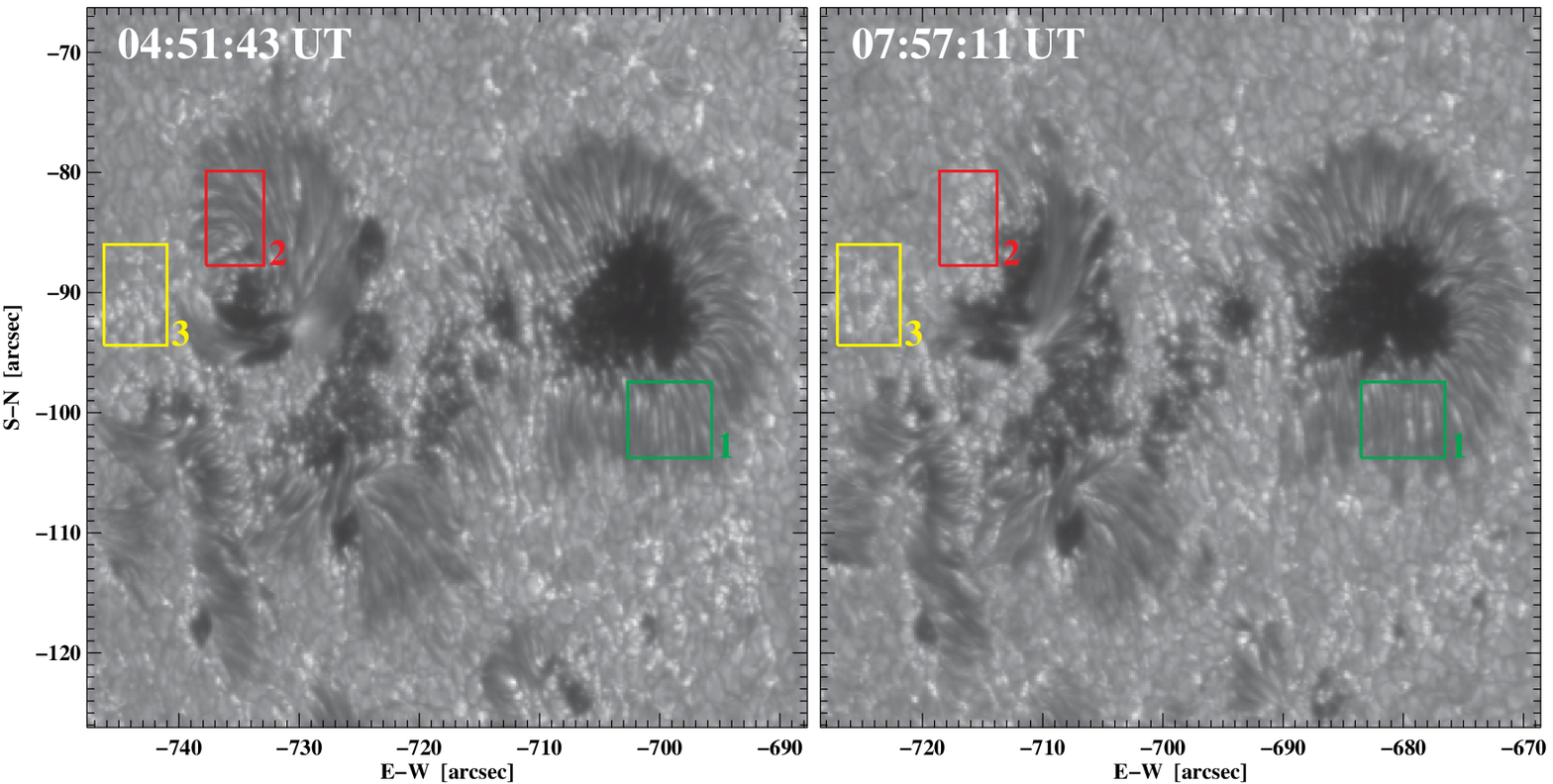}
\caption{Same as Figure 2 but for the 2007 June 4 M8.9 event.}
\end{figure}

\begin{figure}
\epsscale{1.0}
\plotone{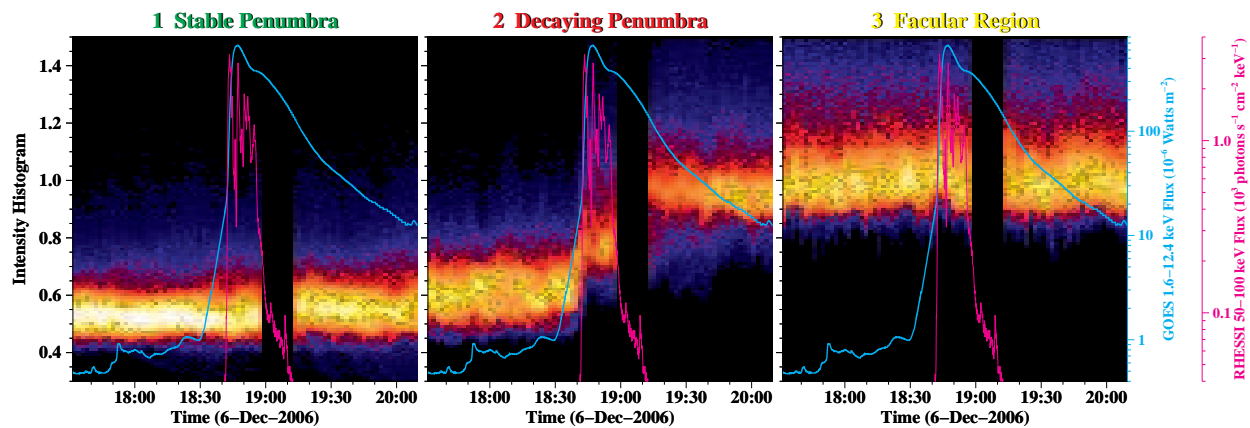}
\caption{The intensity histograms (shown in colors along the Y-axis with the brightness of the color representing the population of pixels falling in the intensity value shown on the Y-axis) as a function of time (X-axis) for the stable penumbra region, decaying penumbra region and facular region (see Figure 2) associated with the X6.5 flare on 2006 December 6. The intensities have been normalized to the quiet Sun intensity. Light curves of RHESSI 50--100~keV hard X-ray (magenta) and GOES 1-8~\AA\ soft X-ray (blue) are overplotted. It is quite evident that the intensity distribution of the decaying penumbra region changes from that of typical penumbra to facular pattern, while the two referential regions do not show detectable, sudden change associated with the flare.}
\end{figure}

\begin{figure}
\epsscale{1.0}
\plotone{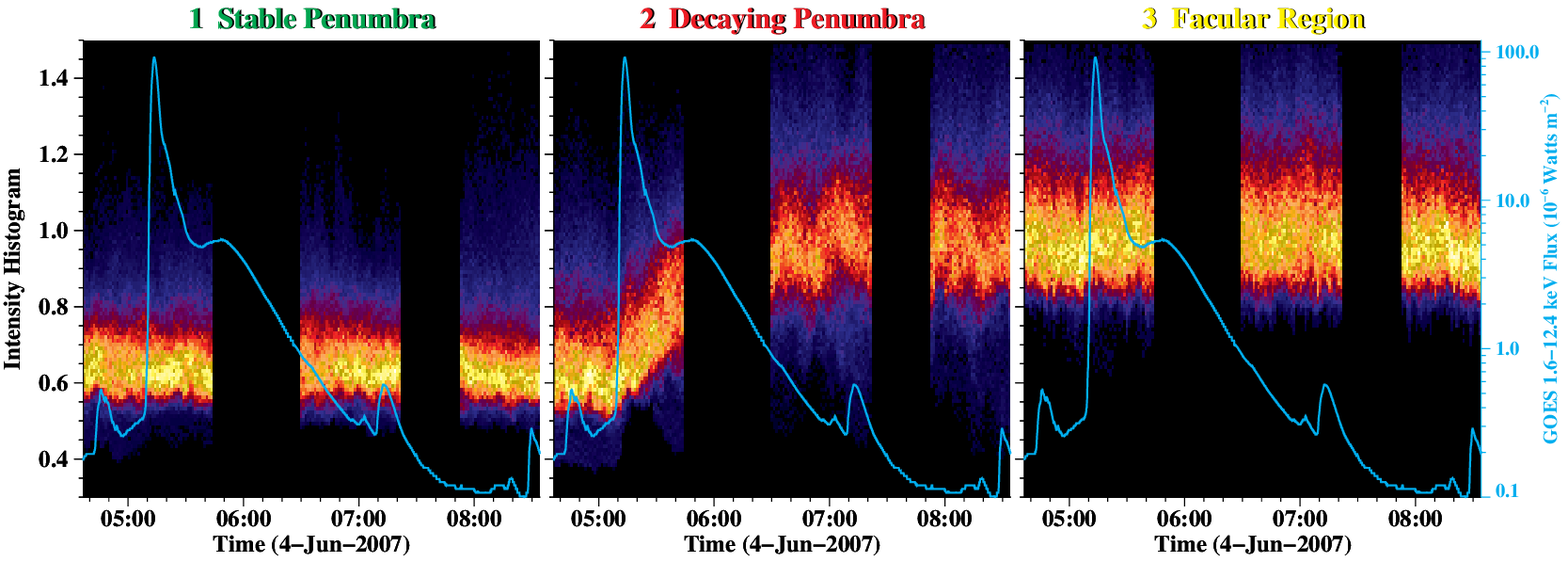}
\caption{Same as Figure 4 but for the 2007 June 4 M8.9 event.}
\end{figure}

\begin{figure}
\epsscale{1.0}
\plotone{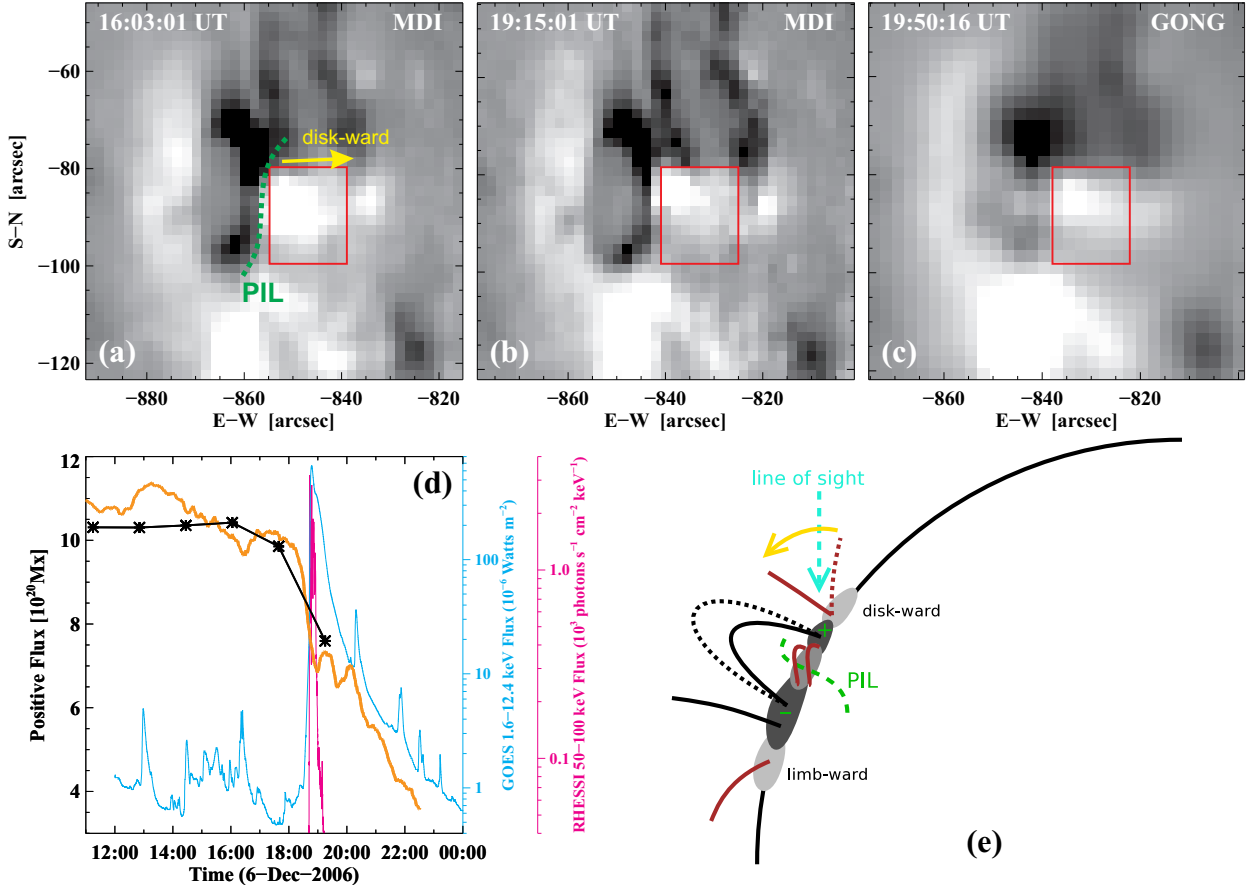}
\caption{(a--c) MDI and GONG LOS magnetograms on 2006 December 6 before and after the X6.5 flare. The image scales for MDI and GONG are 2.0 and 2.4$^{\prime\prime}$~pixel$^{-1}$, respectively. The red box covers the decaying penumbral region. (d) Time evolution of the LOS magnetic flux summed in the red box region. Black points are from MDI magnetograms and orange line is from GONG magnetograms available for this event. The GONG flux is multiplied by a factor of 0.65 in order to be in roughly the same level as MDI flux. (e) Conceptual illustration of the geometry of the $\delta$-spot close to the limb and the change of the field of the decaying penumbra relative to LOS direction (i.e., turning from the red dotted line (horizontal) to the red solid line (vertical)). The decaying penumbra is located at the disk-ward side of the flaring PIL.}
\end{figure}

\begin{figure}
\epsscale{1.0}
\plotone{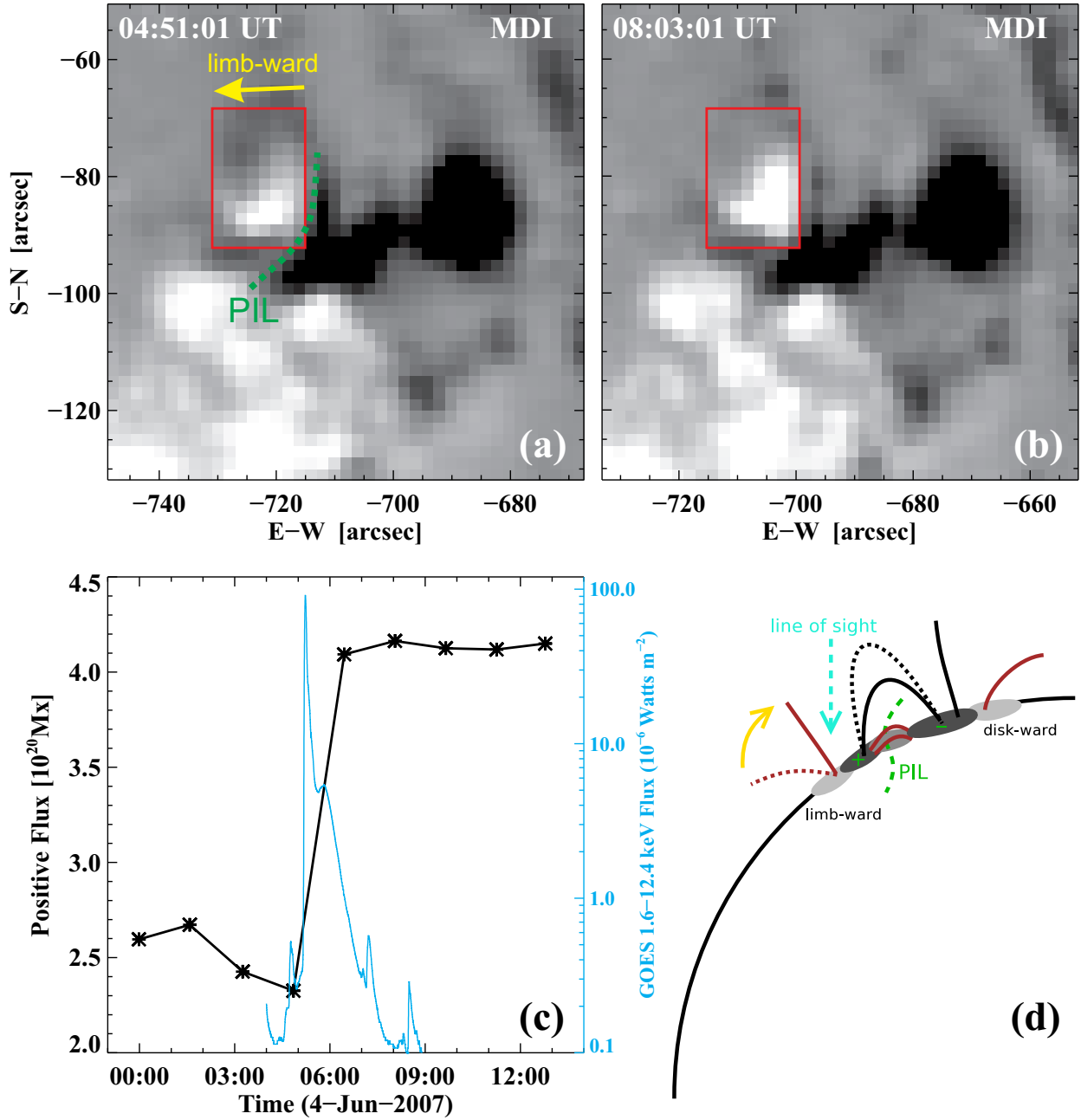}
\caption{Same as Figure 6 but for the 2007 June 4 M8.9 event. MDI magnetograms fully cover this event. The decaying penumbra is located at the limb-ward side of the flaring PIL.}
\end{figure}

\begin{figure}
\epsscale{1.0}
\plotone{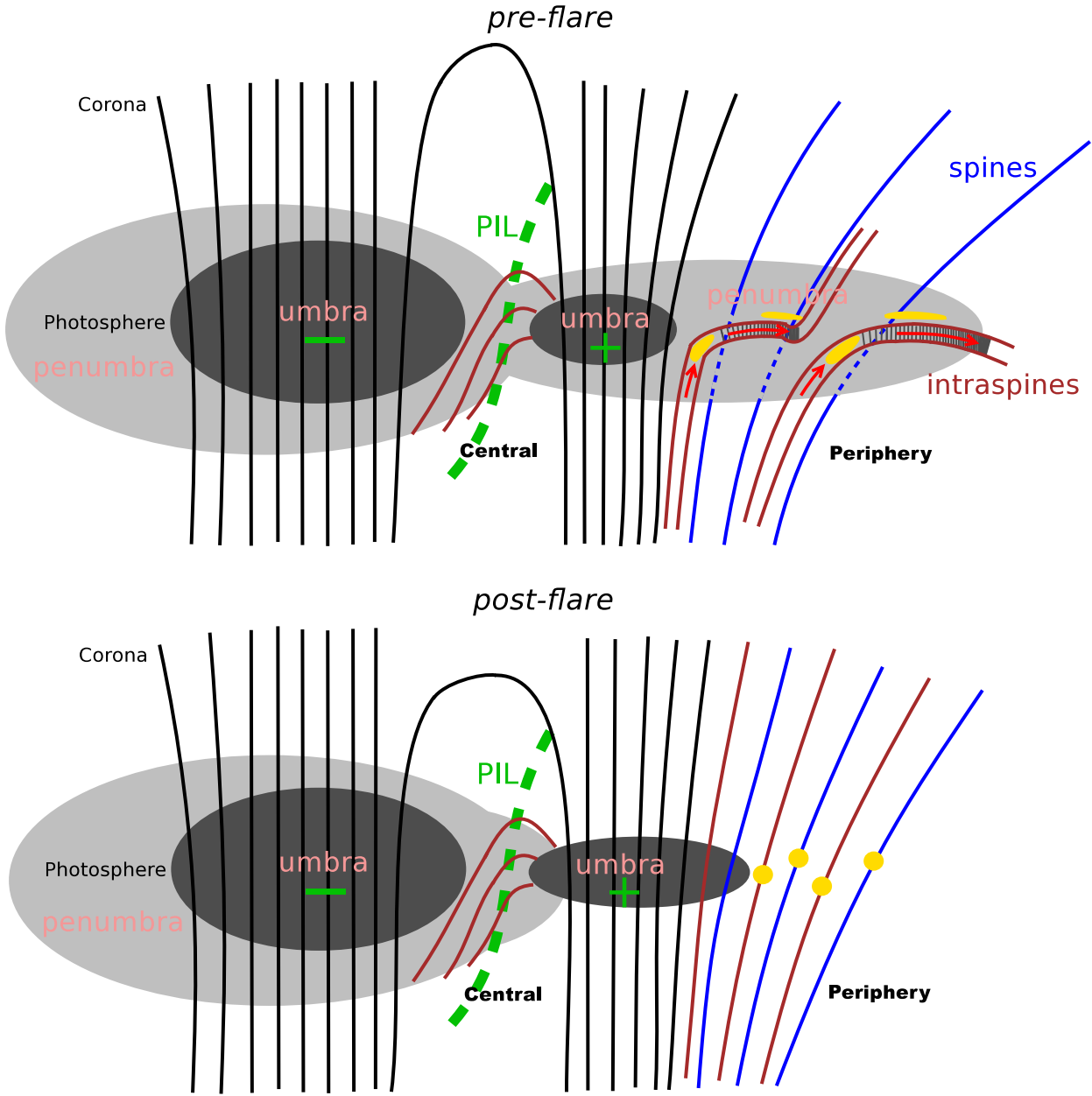}
\caption{Schematic pictures of a typical $\delta$ spot configuration, demonstrating how the uncombed penumbral magnetic fields could be combed by flares leading to the disappearance of the peripheral penumbra. See Section 3 for details.}
\end{figure}


\begin{thebibliography}{57}
\expandafter\ifx\csname natexlab\endcsname\relax\def\natexlab#1{#1}\fi

\bibitem[{{Beck}(2011)}]{BeckC2011A&A...525A.133B}
{Beck}, C. 2011, \aap, 525, A133

\bibitem[{{Bellot Rubio} {et~al.}(2004){Bellot Rubio}, {Balthasar}, \&
  {Collados}}]{BellotRubio+etal2004A&A...427..319B}
{Bellot Rubio}, L.~R., {Balthasar}, H., \& {Collados}, M. 2004, \aap, 427, 319

\bibitem[{{Bellot Rubio} {et~al.}(2003){Bellot Rubio}, {Balthasar}, {Collados},
  \& {Schlichenmaier}}]{BellotRubio+etal2003A&A...403L..47B}
{Bellot Rubio}, L.~R., {Balthasar}, H., {Collados}, M., \& {Schlichenmaier}, R.
  2003, \aap, 403, L47

\bibitem[{{Bellot Rubio} {et~al.}(2006){Bellot Rubio}, {Schlichenmaier}, \&
  {Tritschler}}]{BellotRubio+etal2006A&A...453.1117B}
{Bellot Rubio}, L.~R., {Schlichenmaier}, R., \& {Tritschler}, A. 2006, \aap,
  453, 1117

\bibitem[{{Bellot Rubio} {et~al.}(2008){Bellot Rubio}, {Tritschler}, \&
  {Mart{\'{\i}}nez Pillet}}]{BellotRubio+etal2008ApJ...676..698B}
{Bellot Rubio}, L.~R., {Tritschler}, A., \& {Mart{\'{\i}}nez Pillet}, V. 2008,
  \apj, 676, 698

\bibitem[{{Borrero} \& {Ichimoto}(2011)}]{Borrero+Ichimoto2011LRSP....8....4B}
{Borrero}, J.~M., \& {Ichimoto}, K. 2011, Living Reviews in Solar Physics, 8, 4

\bibitem[{{Borrero} {et~al.}(2005){Borrero}, {Lagg}, {Solanki}, \&
  {Collados}}]{Borrero+etal2005A&A...436..333B}
{Borrero}, J.~M., {Lagg}, A., {Solanki}, S.~K., \& {Collados}, M. 2005, \aap,
  436, 333

\bibitem[{{Borrero} \& {Solanki}(2008)}]{Borrero+Solanki2008ApJ...687..668B}
{Borrero}, J.~M., \& {Solanki}, S.~K. 2008, \apj, 687, 668

\bibitem[{{Borrero} {et~al.}(2006){Borrero}, {Solanki}, {Lagg},
  {Socas-Navarro}, \& {Lites}}]{Borrero+etal2006A&A...450..383B}
{Borrero}, J.~M., {Solanki}, S.~K., {Lagg}, A., {Socas-Navarro}, H., \&
  {Lites}, B. 2006, \aap, 450, 383

\bibitem[{{Chen} {et~al.}(2007){Chen}, {Liu}, {Song}, {Deng}, {Tan}, \&
  {Wang}}]{ChenW+etal2007ChJAA...7..733C}
{Chen}, W., {Liu}, C., {Song}, H., {Deng}, N., {Tan}, C., \& {Wang}, H. 2007,
  Chinese Journal of Astronomy and Astrophysics, 7, 733

\bibitem[{{Deng} {et~al.}(2011){Deng}, {Liu}, {Prasad Choudhary}, \&
  {Wang}}]{DengN+etal2011ApJ...733L..14D}
{Deng}, N., {Liu}, C., {Prasad Choudhary}, D., \& {Wang}, H. 2011, \apjl, 733,
  L14

\bibitem[{{Deng} {et~al.}(2005){Deng}, {Liu}, {Yang}, {Wang}, \&
  {Denker}}]{Deng+etal2005ApJ...623.1195D}
{Deng}, N., {Liu}, C., {Yang}, G., {Wang}, H., \& {Denker}, C. 2005, \apj, 623,
  1195

\bibitem[{{Deng} {et~al.}(2010){Deng}, {Prasad Choudhary}, \&
  {Balasubramaniam}}]{Deng+etal2010ApJ...719..385D}
{Deng}, N., {Prasad Choudhary}, D., \& {Balasubramaniam}, K.~S. 2010, \apj,
  719, 385

\bibitem[{{Fan}(2010)}]{FanY2010ApJ...719..728F}
{Fan}, Y. 2010, \apj, 719, 728

\bibitem[{{Fisher} {et~al.}(2011){Fisher}, {Bercik}, {Welsch}, \&
  {Hudson}}]{Fisher+etal2010arXiv1006.5247F}
{Fisher}, G.~H., {Bercik}, D.~J., {Welsch}, B.~T., \& {Hudson}, H.~S. 2011,
   \solphys, 415

\bibitem[{{Gizon} {et~al.}(2000){Gizon}, {Duvall}, \&
  {Larsen}}]{gizon+duvall+larsen2000}
{Gizon}, L., {Duvall}, Jr., T.~L., \& {Larsen}, R.~M. 2000, Journal of
  Astrophysics and Astronomy, 21, 339

\bibitem[{{Gosain} {et~al.}(2009){Gosain}, {Venkatakrishnan}, \& {Tiwari}}]{Gosain09}
{Gosain}, S., {Venkatakrishnan}, P. \& {Tiwari}, S.~K. 2000, \apjl, 706, L240


\bibitem[{{Hudson}(2000)}]{Hudson2000ApJ...531L..75H}
{Hudson}, H.~S. 2000, \apjl, 531, L75

\bibitem[{{Hudson}(2011)}]{Hudson2011SSRv..158....5H}
---. 2011, \ssr, 158, 5

\bibitem[{{Hudson} {et~al.}(2008){Hudson}, {Fisher}, \&
  {Welsch}}]{Hudson+Fisher+Welsch2008ASPC..383..221H}
{Hudson}, H.~S., {Fisher}, G.~H., \& {Welsch}, B.~T. 2008, in Astronomical
  Society of the Pacific Conference Series, Vol. 383, Subsurface and
  Atmospheric Influences on Solar Activity, ed. {R.~Howe, R.~W.~Komm,
  K.~S.~Balasubramaniam, \& G.~J.~D.~Petrie}, 221--+

\bibitem[{{Ichimoto}(2010)}]{Ichimoto2010mcia.conf..186I}
{Ichimoto}, K. 2010, in Magnetic Coupling between the Interior and Atmosphere
  of the Sun, ed. {S.~S.~Hasan \& R.~J.~Rutten}, 186--192

\bibitem[{{Ichimoto} {et~al.}(2007){Ichimoto}, {Shine}, {Lites}, {Kubo},
  {Shimizu}, {Suematsu}, {Tsuneta}, {Katsukawa}, {Tarbell}, {Title}, {Nagata},
  {Yokoyama}, \& {Shimojo}}]{Ichimoto+etal2007PASJ...59S.593I}
{Ichimoto}, K., {Shine}, R.~A., {Lites}, B., {Kubo}, M., {Shimizu}, T.,
  {Suematsu}, Y., {Tsuneta}, S., {Katsukawa}, Y., {Tarbell}, T.~D., {Title},
  A.~M., {Nagata}, S., {Yokoyama}, T., \& {Shimojo}, M. 2007, \pasj, 59, 593

\bibitem[{{Jur{\v c}{\'a}k} \& {Bellot
  Rubio}(2008)}]{Jurcak+BellotRubio2008A&A...481L..17J}
{Jur{\v c}{\'a}k}, J., \& {Bellot Rubio}, L.~R. 2008, \aap, 481, L17

\bibitem[{{Kosovichev} \&
  {Zharkova}(2001)}]{Kosovichev+Zharkova2001ApJ...550L.105K}
{Kosovichev}, A.~G., \& {Zharkova}, V.~V. 2001, \apjl, 550, L105

\bibitem[{{Kubo} {et~al.}(2011){Kubo}, {Ichimoto}, {Lites}, \&
  {Shine}}]{KuboM+etal2011ApJ...731...84K}
{Kubo}, M., {Ichimoto}, K., {Lites}, B.~W., \& {Shine}, R.~A. 2011, \apj, 731,
  84

\bibitem[{{Kubo} {et~al.}(2008){Kubo}, {Lites}, {Ichimoto}, {Shimizu},
  {Suematsu}, {Katsukawa}, {Tarbell}, {Shine}, {Title}, {Nagata}, \&
  {Tsuneta}}]{KuboM+etal2008ApJ...681.1677K}
{Kubo}, M., {Lites}, B.~W., {Ichimoto}, K., {Shimizu}, T., {Suematsu}, Y.,
  {Katsukawa}, Y., {Tarbell}, T.~D., {Shine}, R.~A., {Title}, A.~M., {Nagata},
  S., \& {Tsuneta}, S. 2008, \apj, 681, 1677

\bibitem[{{Langhans} {et~al.}(2005){Langhans}, {Scharmer}, {Kiselman},
  {L{\"o}fdahl}, \& {Berger}}]{Langhans+etal2005A&A...436.1087L}
{Langhans}, K., {Scharmer}, G.~B., {Kiselman}, D., {L{\"o}fdahl}, M.~G., \&
  {Berger}, T.~E. 2005, \aap, 436, 1087

\bibitem [{Li} {et~al.} (2009) {Li} \& {Zhang}] { }
 {Li}, L., \& {Zhang}, J., 2009, \apjl, 706, L17

\bibitem[{{Li} {et~al.}(2011){Li}, {Jing}, {Fan}, \&
  {Wang}}]{LiY+etal2011ApJ...727L..19L}
{Li}, Y., {Jing}, J., {Fan}, Y., \& {Wang}, H. 2011, \apjl, 727, L19

\bibitem[{{Liggett} \& {Zirin}(1983)}]{Liggett+Zirin1983SoPh...84....3L}
{Liggett}, M., \& {Zirin}, H. 1983, \solphys, 84, 3

\bibitem[{{Lites} {et~al.}(1993){Lites}, {Elmore}, {Seagraves}, \&
  {Skumanich}}]{lites+etal1993}
{Lites}, B.~W., {Elmore}, D.~F., {Seagraves}, P., \& {Skumanich}, A.~P. 1993,
  \apj, 418, 928

\bibitem[{{Lites} {et~al.}(2008){Lites}, {Kubo}, {Socas-Navarro}, {Berger},
  {Frank}, {Shine}, {Tarbell}, {Title}, {Ichimoto}, {Katsukawa}, {Tsuneta},
  {Suematsu}, {Shimizu}, \& {Nagata}}]{Lites+etal2008ApJ...672.1237L}
{Lites}, B.~W., {Kubo}, M., {Socas-Navarro}, H., {Berger}, T., {Frank}, Z.,
  {Shine}, R., {Tarbell}, T., {Title}, A., {Ichimoto}, K., {Katsukawa}, Y.,
  {Tsuneta}, S., {Suematsu}, Y., {Shimizu}, T., \& {Nagata}, S. 2008, \apj,
  672, 1237

\bibitem[{{Liu} {et~al.}(2005){Liu}, {Deng}, {Liu}, {Falconer}, {Goode},
  {Denker}, \& {Wang}}]{LiuC+etal2005ApJ...622..722L}
{Liu}, C., {Deng}, N., {Liu}, Y., {Falconer}, D., {Goode}, P.~R., {Denker}, C.,
  \& {Wang}, H. 2005, \apj, 622, 722

\bibitem[{{Liu} \& {Wang}(2009)}]{LiuR+WangH2009ApJ...703L..23L}
{Liu}, R., \& {Wang}, H. 2009, \apjl, 703, L23

\bibitem[{{Mattews} {et~al.}(2011){Mathews}, {Zharkov}, \&
  {Zharkova}}]{Mattews11}
{Matthews}, S.~A., {Zharkov}, S. \& {Zharkova}, V.~V. 2011, \apj, 739, 71


\bibitem[{{Petrie} \& {Sudol}(2010)}]{Petrie+Sudol2010ApJ...724.1218P}
{Petrie}, G.~J.~D., \& {Sudol}, J.~J. 2010, \apj, 724, 1218

\bibitem[{{Rempel}(2011{\natexlab{a}})}]{Rempel2011ApJ...729....5R}
{Rempel}, M. 2011{\natexlab{a}}, \apj, 729, 5

\bibitem[{{Rempel} {et~al.}(2009){Rempel}, {Sch{\"u}ssler}, {Cameron}, \&
  {Kn{\"o}lker}}]{Rempel+etal2009Sci...325..171R}
{Rempel}, M., {Sch{\"u}ssler}, M., {Cameron}, R.~H., \& {Kn{\"o}lker}, M. 2009,
  Science, 325, 171

\bibitem[{{Rempel}(2011{\natexlab{b}})}]{RempelM2011SPD....42.1001R}
{Rempel}, M.~D. 2011{\natexlab{b}}, in AAS/Solar Physics Division Abstracts
  \#42, 1001

\bibitem[{{Rimmele} \& {Marino}(2006)}]{rimmele+marino2006}
{Rimmele}, T., \& {Marino}, J. 2006, \apj, 646, 593

\bibitem[{{Scharmer} \& {Spruit}(2006)}]{Scharmer+Spruit2006A&A...460..605S}
{Scharmer}, G.~B., \& {Spruit}, H.~C. 2006, \aap, 460, 605

\bibitem[{{Schlichenmaier} {et~al.}(2005){Schlichenmaier}, {Bellot Rubio}, \&
  {Tritschler}}]{Schlichenmaier+etal2005AN....326..301S}
{Schlichenmaier}, R., {Bellot Rubio}, L.~R., \& {Tritschler}, A. 2005,
  Astronomische Nachrichten, 326, 301

\bibitem[{{Schlichenmaier} {et~al.}(1998){Schlichenmaier}, {Jahn}, \&
  {Schmidt}}]{schlichenmaier+jahn+schmidt1998b}
{Schlichenmaier}, R., {Jahn}, K., \& {Schmidt}, H.~U. 1998, \aap, 337, 897

\bibitem[{{Solanki} \& {Montavon}(1993)}]{Solanki+Montavon1993A&A...275..283S}
{Solanki}, S.~K., \& {Montavon}, C.~A.~P. 1993, \aap, 275, 283

\bibitem[{{Spirock} {et~al.}(2002){Spirock}, {Yurchyshyn}, \&
  {Wang}}]{Spirock+etal2002ApJ...572.1072S}
{Spirock}, T.~J., {Yurchyshyn}, V.~B., \& {Wang}, H. 2002, \apj, 572, 1072

\bibitem[{{Spruit}(1976)}]{spruit1976}
{Spruit}, H.~C. 1976, \solphys, 50, 269

\bibitem[{{Spruit} \& {Scharmer}(2006)}]{Spruit+Scharmer2006A&A...447..343S}
{Spruit}, H.~C., \& {Scharmer}, G.~B. 2006, \aap, 447, 343

\bibitem[{{Stanchfield} {et~al.}(1997){Stanchfield}, {Thomas}, \&
  {Lites}}]{stanchfield+thomas+lites1997}
{Stanchfield}, D.~C.~H., {Thomas}, J.~H., \& {Lites}, B.~W. 1997, \apj, 477,
  485

\bibitem[{{Sudol} \& {Harvey}(2005)}]{Sudol+Harvey2005ApJ...635..647S}
{Sudol}, J.~J., \& {Harvey}, J.~W. 2005, \apj, 635, 647

\bibitem[{{Tan} {et~al.}(2009){Tan}, {Chen}, {Abramenko}, \&
  {Wang}}]{Tan+etal2009ApJ...690.1820T}
{Tan}, C., {Chen}, P.~F., {Abramenko}, V., \& {Wang}, H. 2009, \apj, 690, 1820

\bibitem[{{Tsuneta} {et~al.}(2008){Tsuneta}, {Ichimoto}, {Katsukawa}, {Nagata},
  {Otsubo}, {Shimizu}, {Suematsu}, {Nakagiri}, {Noguchi}, {Tarbell}, {Title},
  {Shine}, {Rosenberg}, {Hoffmann}, {Jurcevich}, {Kushner}, {Levay}, {Lites},
  {Elmore}, {Matsushita}, {Kawaguchi}, {Saito}, {Mikami}, {Hill}, \&
  {Owens}}]{Tsuneta+etal2008SoPh..249..167T}
{Tsuneta}, S., {Ichimoto}, K., {Katsukawa}, Y., {Nagata}, S., {Otsubo}, M.,
  {Shimizu}, T., {Suematsu}, Y., {Nakagiri}, M., {Noguchi}, M., {Tarbell}, T.,
  {Title}, A., {Shine}, R., {Rosenberg}, W., {Hoffmann}, C., {Jurcevich}, B.,
  {Kushner}, G., {Levay}, M., {Lites}, B., {Elmore}, D., {Matsushita}, T.,
  {Kawaguchi}, N., {Saito}, H., {Mikami}, I., {Hill}, L.~D., \& {Owens}, J.~K.
  2008, \solphys, 249, 167

\bibitem[{{Wang}(1992)}]{WangH1992SoPh..140...85W}
{Wang}, H. 1992, \solphys, 140, 85

\bibitem[{{Wang}(2006)}]{WangH2006ApJ...649..490W}
---. 2006, \apj, 649, 490

\bibitem[{{Wang} {et~al.}(1994){Wang}, {Ewell}, {Zirin}, \&
  {Ai}}]{Wang+etal1994ApJ...424..436W}
{Wang}, H., {Ewell}, Jr., M.~W., {Zirin}, H., \& {Ai}, G. 1994, \apj, 424, 436

\bibitem[{{Wang} \& {Liu}(2010)}]{Wang+Liu2010ApJ...716L.195W}
{Wang}, H., \& {Liu}, C. 2010, \apjl, 716, L195

\bibitem[{{Wang} {et~al.}(2004{\natexlab{a}}){Wang}, {Liu}, {Qiu}, {Deng},
  {Goode}, \& {Denker}}]{WangH+etal2004ApJ...601L.195W}
{Wang}, H., {Liu}, C., {Qiu}, J., {Deng}, N., {Goode}, P.~R., \& {Denker}, C.
  2004{\natexlab{a}}, \apjl, 601, L195

\bibitem[{{Wang} {et~al.}(2004{\natexlab{b}}){Wang}, {Qiu}, {Jing}, {Spirock},
  {Yurchyshyn}, {Abramenko}, {Ji}, \& {Goode}}]{Wang+etal2004ApJ...605..931W}
{Wang}, H., {Qiu}, J., {Jing}, J., {Spirock}, T.~J., {Yurchyshyn}, V.,
  {Abramenko}, V., {Ji}, H., \& {Goode}, P.~R. 2004{\natexlab{b}}, \apj, 605,
  931

\bibitem[{{Wang} {et~al.}(2002){Wang}, {Spirock}, {Qiu}, {Ji}, {Yurchyshyn},
  {Moon}, {Denker}, \& {Goode}}]{Wang+etal2002ApJ...576..497W}
{Wang}, H., {Spirock}, T.~J., {Qiu}, J., {Ji}, H., {Yurchyshyn}, V., {Moon},
  Y.-J., {Denker}, C., \& {Goode}, P.~R. 2002, \apj, 576, 497

\bibitem[{{Wang} {et~al.}(2012){Wang}, {Liu}, {Liu}, {Deng}, {Liu}, \&
  {Wang}}]{WangS+etal2011arXiv1103.0027W}
{Wang}, S., {Liu}, C., {Liu}, R., {Deng}, N., {Liu}, Y., \& {Wang}, H. 2012,
  \apjl, accepted

\bibitem[{{Yurchyshyn} {et~al.}(2004){Yurchyshyn}, {Wang}, {Abramenko},
  {Spirock}, \& {Krucker}}]{Yurchyshyn+etal2004ApJ...605..546Y}
{Yurchyshyn}, V., {Wang}, H., {Abramenko}, V., {Spirock}, T.~J., \& {Krucker},
  S. 2004, \apj, 605, 546

\end{thebibliography}
\end{document}